\documentclass{pasj00}
\usepackage{natbib}
\draft
\usepackage{setspace}
\begin{document}
\SetRunningHead{Irina N. Kitiashvili}{Spectro-Polarimetric Properties of Small-Scale Plasma Eruptions}

\title{Spectro-Polarimetric Properties of Small-Scale\\Plasma Eruptions Driven by Magnetic Vortex Tubes}

\author{Irina~N. \textsc{Kitiashvili} %
  \thanks{Stanford, CA 94305, USA}}
\affil{Stanford University}
\email{inkitiashvili@gmail.com}

%


\maketitle

\begin{abstract}
Highly turbulent nature of convection on the Sun causes strong multi-scale interaction of subsurface layers with the photosphere and chromosphere. According to realistic 3D radiative MHD numerical simulations ubiquitous small-scale vortex tubes are generated by turbulent flows below the visible surface and concentrated in the intergranular lanes. The vortex tubes can capture and amplify magnetic field, penetrate into chromospheric layers and initiate quasi-periodic flow eruptions that generates Alfv\'enic waves, transport mass and energy into the solar atmosphere. The simulations revealed high-speed flow patterns, and complicated thermodynamic and magnetic structures in the erupting vortex tubes. The spontaneous eruptions are initiated and driven by strong pressure gradients in the near-surface layers, and accelerated by the Lorentz force in the low chromosphere. In this paper, the simulation data are used to further investigate the dynamics of the eruptions, their spectro-polarimetric characteristics for the Fe I 6301.5 and 6302.5~\AA~spectral lines, and demonstrate expected signatures of the eruptions in the Hinode SP data. We found that the complex dynamical structure of vortex tubes (downflows in the vortex core and upflows on periphery) can be captured by the Stokes I profiles. During an eruption, the ratio of down and upflows can suddenly change, and this effect can be observed in the Stokes~V profile. Also, during the eruption the linear polarization signal increases, and this also can be detected with Hinode SP.
\end{abstract}

\section{Introduction}
The fast growth of observational capabilities from space (e.g., Hinode, SDO) and ground-based (e.g., SST, NST/BBSO) telescopes opens new opportunities to investigate  highly dynamical behavior of magnetized plasma flows in the solar atmosphere across all resolved scales. In particular, the eruptive dynamics of the Sun is represented from small subarcsecond scales in the form of tiny intergranular jets \citep{Yurchyshyn2011} to large scales in the form of coronal mass ejections and flares \citep{Shibata2011,Webb2012}. The origin of these eruptions is still unclear. While the large-scale eruptions can be due to instabilities and magnetic reconnection in the higher atmosphere, the small-scale eruptions are clearly linked to turbulent convection. Their understanding requires very detailed analysis of high-resolution multiwavelength data and accurate modeling of the magnetized turbulent plasma of the solar subsurface and atmosphere. Numerical radiative MHD simulations, based on '{\it ab initio}' physical principles, have been able to reproduce observed phenomena with a high degree of realism, allowed us to understand their physics and also predict new effects that are difficult to detect in observations \cite[e.g][]{Stein1998,Vogler2005,Jacoutot2008a,Kitiashvili2010,Kitiashvili2013,Gudiksen2011}. However, dramatic differences between the physical conditions of the convective zone and the corona (where the MHD approximation is not valid), restrict extention of the numerical MHD models into the high atmospheric layers. Previous simulation results showed that turbulent vortex tubes generated near the surface by convective and the Kelvin-Helmholtz instabilities \citep{Kitiashvili2012} can trigger various dynamical processes deeper in the interior and higher in the atmosphere, such as acoustic waves excitation, spontaneous formation of compact pore-like magnetic structures and other phenomena \citep[e.g.][]{Kitiashvili2010,Kitiashvili2011}.

In this paper, I present results of spectro-polarimetric modeling of the radiative MHD simulations  \cite{Kitiashvili2013a} that shed light on the mechanism of small-scale eruptions in the solar atmosphere, linked to the dynamics of turbulent magnetized vortex tubes. The results are focused on small-scale flow dynamics from the subsurface layers to 1~Mm  above the photosphere, and demonstrate that spontaneously initiated quasi-periodic upflows associated with the vortex tube dynamics, in the presence of magnetic field, produce small-scale jet-like ejections of plasma generating shocks in the atmosphere. The physical properties of these eruptions, and their implication for high-resolution observations are discussed. The small-scale vortex tubes, mostly concentrated in the intergranular lanes, are hard to detect in broad-band images, however, their dynamics leads to very specific signatures in spectro-polarimetric Stokes profiles.

\section{Physical properties of spontaneous small-scale flow eruptions}
Simulations of the turbulent solar convection have shown generation of ubiquitous small-scale vortices resulting from convective overturning and Kelvin-Helmholtz-type (shear) instabilities in the highly stratified medium \citep{Kitiashvili2012}. Such vortical structures were found in various numerical simulations of the solar-type convection \citep[e.g.][]{Zirker1993,Stein1998,Kitiashvili2010,Kitiashvili2011,Steiner2010} and observations \citep[e.g.][]{Stenflo1975,Wang1995,Bonet2008,Bonet2010,Wedemeyer-Bohm2009}. According to the simulation results these vortical structures are characterized by sharp decreasing of temperature, density and gas pressure, strong, $\sim 7-8$~kms/s in the vortex core, downflows, and often supersonic horizontal flows around the vortex cores. The vortex tubes, which are usually located in the intergranular lanes have life-time much longer then the solar granules (up to 1~hour).

\begin{figure}
 \begin{center}
  \includegraphics[width=16.5cm]{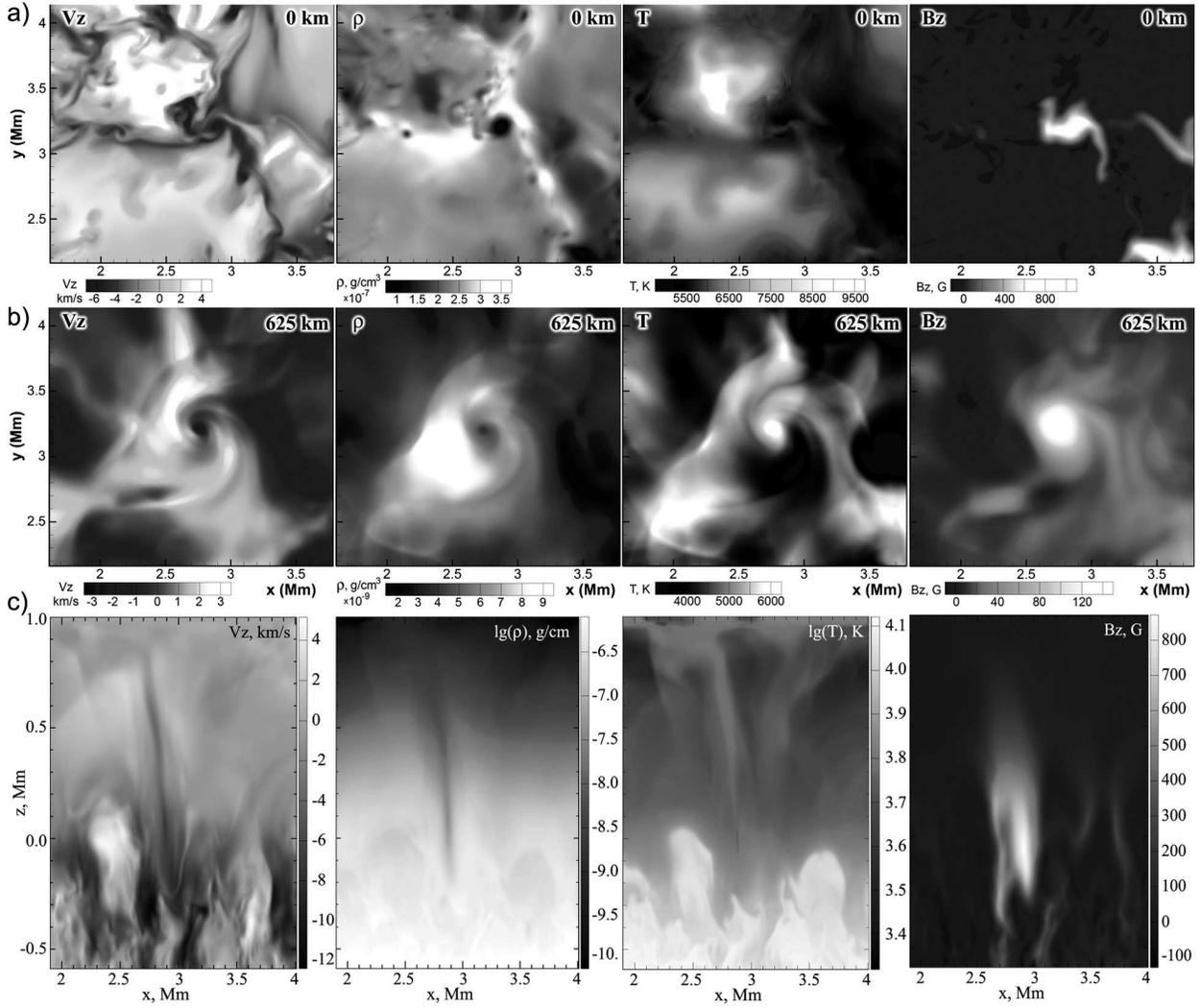}
 \end{center}
\caption{Snapshots of a fraction of box domain in photosphere (panel $a$) and 625~km above solar surface (panel $b$) and vertical cuts through the vortex tube (panel $c$) for different quantities (from left to right): vertical velocity, density, temperature and vertical magnetic field.}\label{fig:2layers_xy}
\end{figure}

Recent simulations revealed that the vortex tubes can penetrate into the chromosphere strongly affecting its thermodynamic properties \citep{Kitiashvili2012a,Kitiashvili2013a}. An example of an erupting vortex tube in Figure~\ref{fig:2layers_xy} shows that in the vortex core the photospheric temperature is lower than the mean temperature, whereas in the chromospheric layers, $\sim 600$~km above the photosphere, the temperature is higher. The typical size of the vortex core is $\sim 100$~km in the photosphere, and about 1~Mm in the chromosphere, 200 -- 300~km is the vortex core size. Such chromospheric helical structures were recently observed by \cite{Wedemeyer-Bohm2009}.

\begin{figure}
 \begin{center}
  \includegraphics[width=15.5cm]{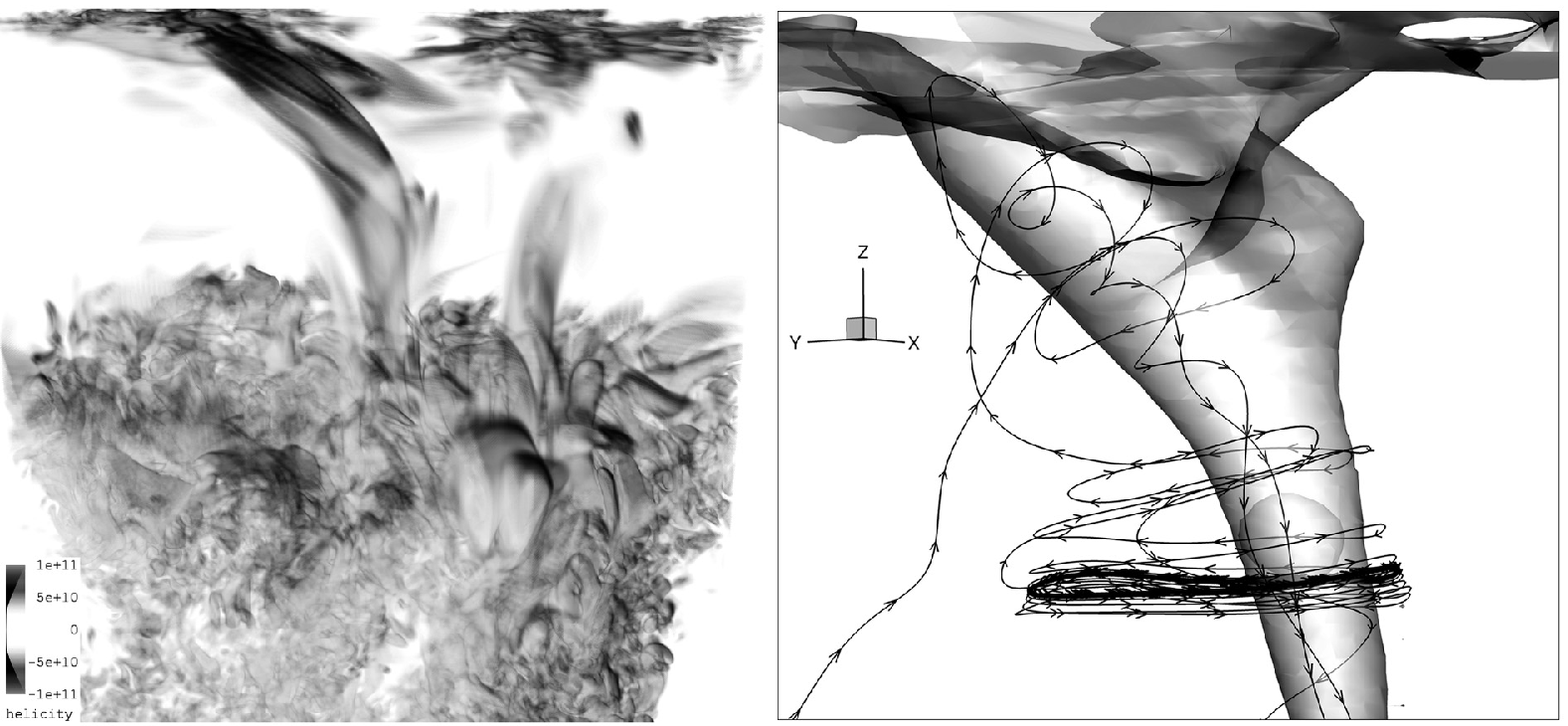}
 \end{center}
\caption{$a$) 3D rendering of the kinetic helicity during a vortex tube eruption; $b$) streamlines illustrating the structure of the velocity field in the vicinity of the vortex core. The vortex core is shown as an isosurface of temperature. In panel $a$ the horizontal and vertical size of the snapshot is about 1.7~Mm}\label{fig:hel-erup}
\end{figure}

Vortex-tube penetration upward into the solar atmosphere is often quasi-periodic, with a period in the range of 2 -- 5~min and accompanied by spontaneous flow ejections \cite{Kitiashvili2013a} and generation of chromospheric shocks. Figure~\ref{fig:2layers_xy} illustrates the complicated structure and dynamics of the eruptions, with downflows in the vortex core and upward eruption flows in the surrounding region. As seen in the temperature distribution, these eruptions are hotter than the surrounding plasma and can provide extra chromospheric heating in addition to the heating through the vortex core (Fig.~\ref{fig:2layers_xy}).

Due to the turbulent nature of the helical motions along the vortex and its interaction with the surface and atmospheric layers, the structure of the eruption is highly inhomogeneous. Figure~\ref{fig:hel-erup}$a$ shows an example of 3D rendering of the kinetic helicity in a small fraction of the computational domain during the eruption, where darker colors correspond to stronger helicity of the flows and demonstrate clustering of the flows. On scale of the whole vortex tube, up- and downflows take place at the same time. Figure~\ref{fig:hel-erup}$b$ demonstrates the topological structure of plasma during the eruption, where the isosurface corresponds to constant temperature, 5800~K, the streamlines represent a sample of the velocity field, illustrating a coexistence of plasma ejection with accompanied by highly twisted flows on the vortex periphery and helical downflows in the vortex core. Thus, the flow structure during the eruption phase remains twisted: the material around the vortex core moves up from the subsurface and near surface layers, and also towards the vortex core from the surrounding region. The plasma is moved up by the twisting flows into the higher atmospheric layers, and, at the same time, in the lower layers the plasma flows down though the vortex core (Fig.~\ref{fig:hel-erup}$b$). These eruptions are initiated in a near-surface 120~km deep layer by impulsive strong pressure gradients and accelerated by the Lorentz force in the atmosphere, where magnetic field effects are dominant \citep{Kitiashvili2013a}.

Because the small-scale of these eruptions it is a challenging task to detect them in observations. In the next section I consider the evolution of Stokes profiles of the Fe~I~6302.5~\AA~line during flow eruptions, and discuss the possibility for detecting these event with Hinode spacecraft.

\begin{figure}
 \begin{center}
  \includegraphics[width=16.5cm]{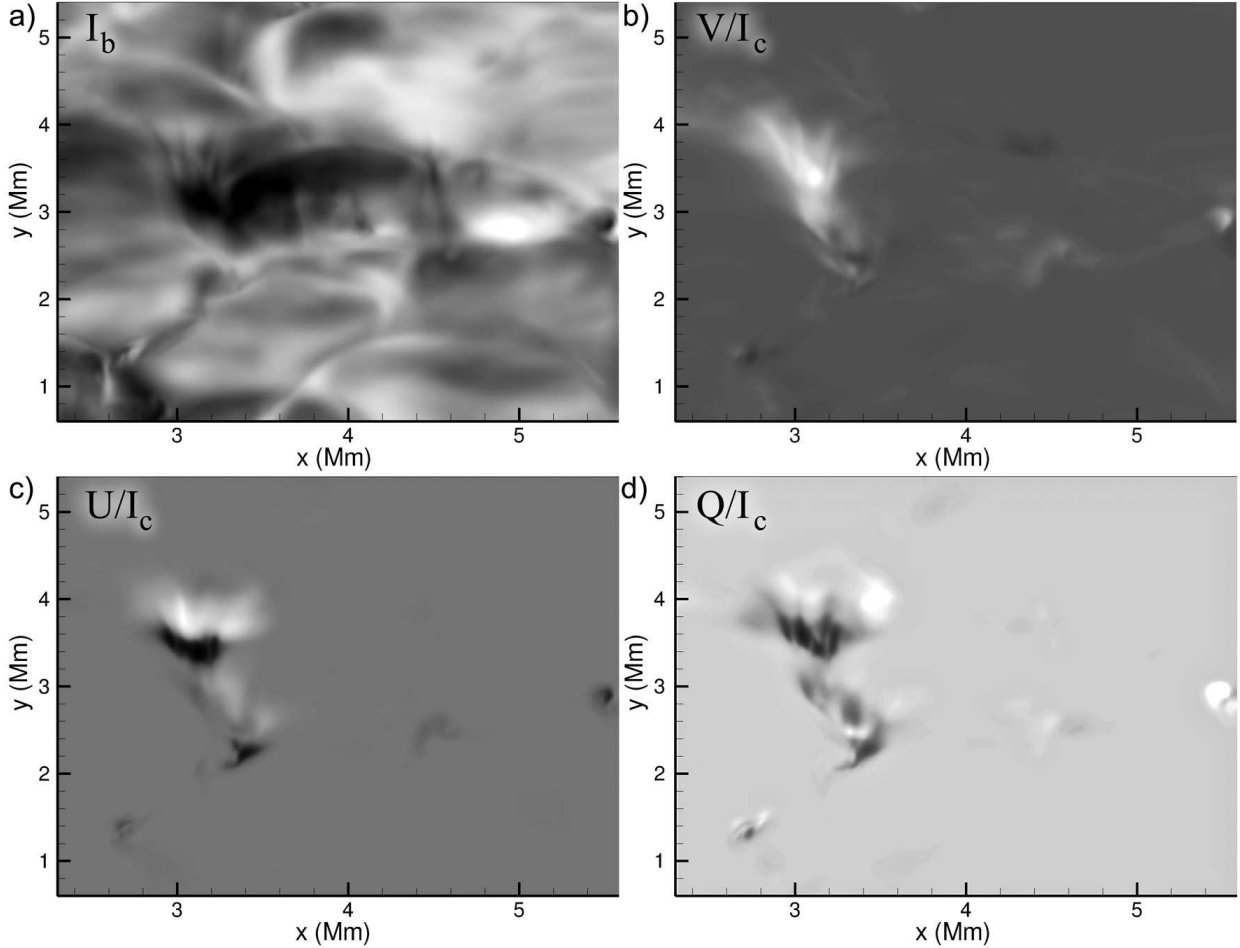}
 \end{center}
\caption{Ultra-fine loops and small-scale eruption as seen in the blue-wing Stokes images of line 6302.5~\AA, calculated from the simulation data: blue shift of Stokes~I (panel $a$), normalized Stokes vectors: V/I${_{\rm c}}$ (panel $b$), U/I${_{\rm c}}$ (panel $c$), Q/I${_{\rm d}}$ (panel $b$).}\label{fig:stokes}
\end{figure}

\section{Spectro-polarimetric properties of the eruptions}
Our simulation results show that the small-scale flow eruptions are ubiquitous in the quiet Sun regions with weak mean magnetic field (e.g. in coronal holes and plage regions). However, locally, in the vortex tubes magnetic field strength can reach $\sim 1$~kG. In fact, it was found that in the strongest magnetic field patches are associated with the strongest flow ejections. Studying the small-scale phenomena that energize the solar atmosphere observationally is a very challenging problem because of requirement of high spatial (subarcsec) and temporal (few seconds) resolutions. The first step to address this problem is to convert our numerical model into spectropolarimetric observables, in particular, the Stokes profiles for the line $\lambda_0=6302.5$~\AA~observed by Hinode. The Stokes profiles are calculated using the SPINOR code \citep{Frutiger2000}.

Figure~\ref{fig:stokes} shows the images of the intensity in the blue wing ($\triangle \lambda = -0.06$~\AA, panel $a$) of this line, and the corresponding Stokes V, U and Q images normalized by continuum intensity (panels $c-d$) for a small fraction of the box domain, assuming that this region is located 60 degrees from the solar disk center (the bottom of images is closer to the disk center). The intensity blue wing (panel $a$) shows numerous structures around granules, some of them represents tiny loops, which are organized in clusters associated with magnetic field patches (panels $a$ and $b$). The signal of transverse magnetic fields, represented by Stokes U and Q, is much weaker than the line-of-sight component signal (Stokes V). In Figure~\ref{fig:stokes}, the gray scale for Stokes U and Q (panels $c$ and $d$) is adjusted to show the fine substructures.

\begin{figure}
 \begin{center}
  \includegraphics[width=16.cm]{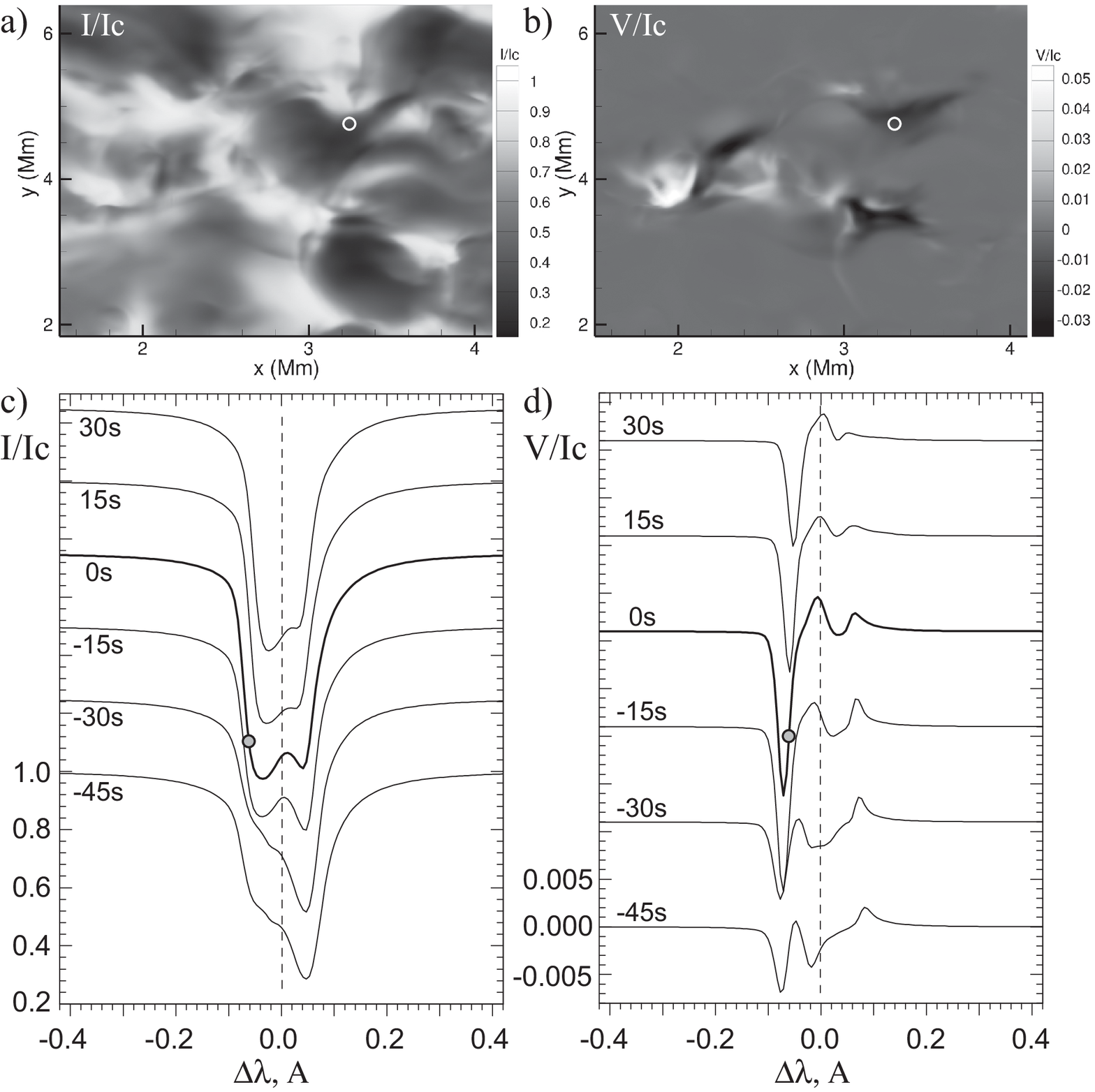}
 \end{center}
\caption{Flow eruption reflected in Stokes I and V. Panels $a$ and $b$ show zoomed images of I/Ic and V/Ic for $\delta \lambda = -0.61$~\AA ($\lambda_0=6401.5$~\AA). Panels $c$ and $d$ illustrate the evolution of Stokes profiles I/Ic and V/Ic. Thick-line profiles and circles on the profiles indicate the location of the slit. Circles on images indicates location of the sample during the flow eruption.}\label{fig:jet}
\end{figure}

In the vicinity of the vortex tubes, associated with the flow eruption, the Stokes profiles have a complicated structure. In particular, the Stokes~I profile usually has two minima (Fig.~\ref{fig:jet}$c$). This line splitting can reflect coexisting down and upflows and creates difficulties for determination of the Doppler shift using the standard line-shift methods. According to our numerical simulations (Fig.~\ref{fig:hel-erup}$b$), the downflows occupy the vortex core, and the upflows occur at the periphery of the vortex tube \citep{Kitiashvili2013}. The coexisting upflow and downflow regions along the line-of-sight make the line profile more complicated that it is usually assumed in observational data analysis. In fact, the Stokes I profile can have even more wiggles, which, are probably, related to shock waves in the chromosphere crossing the line-of-sight, or to the complex structuring of the vortex environment due to the turbulent swirling flows. Such fine-scale spectral dynamics requires additional studies.

Figure~\ref{fig:jet} shows a snapshot of a portion of our box domain, where `observation' was taken in the blue wing of line $\lambda_0=6301.5$~\AA~at $\delta\lambda=-0.06$~\AA. The blue-shifted structure in the middle of the snapshot (Stokes I/Ic, panel $a$) clearly correlates with the distribution of the  line-of-sight magnetic field (Stokes V/Ic, panel $b$). Such coupling of the flow dynamics and magnetic fields, which probably can be detected with SOT/Hinode and the future Solar-C mission, is illustrated by the time-evolution of the Stokes profiles in panels $c$ and $d$. The time-difference between the Stokes profiles is 15~s, the initial moment of time is arbitrary and is set to 0~sec for the snapshots shown in panels $a$ and $b$. Small circles indicate the location of the Stokes samples during the eruption. In the region of the flow eruption the intensity profiles are deformed due to the vortex tube flow structure, when the contribution of down and upflow `components' change at different stages of the eruption (Fig.~\ref{fig:jet} $a$). Such flow dynamics causes `running-wave' variations across the Stokes profiles. During the eruptive phase the polarization of light increases due to the locally amplified magnetic field (Fig.~\ref{fig:jet} $b$); this effect makes it possible to detect the eruptions in observations with relatively low spatial resolution. The Stokes U and Q profiles which describes the transverse components of magnetic field have essentially more complicated structure than Stokes V, and also shows fast changes of the profiles. However, the amplitude of the fluctuations probably is too small to be detectable with SOT/Hinode.

\section{Discussion and conclusion}
Growing numerical and observational capabilities allow us to detect and investigate very interesting fine-scale MHD phenomena causing intense and very dynamic interactions between the surface layers and the chromosphere in quiet-Sun regions. It becomes more and more evident that the small-scale of vortex tube dynamics can play significant role in the various observed phenomena, such as formation of pore-like structures, acoustic waves excitation, quasi-periodic flow eruptions, and substantially contribute to the energy exchange between the turbulent surface and subsurface layers and the chromosphere \citep{Kitiashvili2010,Kitiashvili2011,Kitiashvili2012a,Kitiashvili2013a,Moll2011,Shelyag2011}. Modern multiwavelength observations confirm the existence of dynamical links between these layers through the vortex tubes \citep{Wedemeyer-Bohm2012}. According to our numerical simulations, spontaneous and ubiquitous flow eruptions occur everywhere in the quiet Sun. Initially these eruptions are driven by impulsive increasing of the pressure gradient due to turbulent motions associated with the vortex tubes. In the chromosphere, where magnetic field effects become dominant, the flows are further accelerated by the Lorentz force of the magnetic field twisted and amplified in the vortex tubes. The simulations predict that the small-scale eruptions are more prominent in the regions with distributed magnetic fields.

The small-scale dynamics is not easy observed in broad-band images even with high spatial resolution. However, these phenomena have very distinct signatures in the Stokes profiles of solar spectral lines. This allows us to develop new approaches for studying the small-scale dynamics in observations. In this paper I used a 3D radiative MHD model with B$_{z_0}$=10~G mean magnetic field \citep{Kitiashvili2013a} to calculate two Hinode lines: 6301.5~\AA~and 6302.5~\AA. The synthetic Stokes profiles for $\lambda_0=6302.5$~\AA~(shown for the line blue wing in Figure~\ref{fig:stokes}) demonstrate the existence of small-scale jets. The Stokes profiles reveal interesting variations in the form of `running-wave' perturbations across the line with time, which reflect the bidirectional flow dynamics of the eruptions. The Stokes images show ubiquitous small-scale elongated structures, which originate in the photosphere and some of them can be identified as ultrafine loops above the granulation. Also, several of these superfine structures can occupy relatively small areas ($100-200$~km in size), which are associated with magnetic field concentrations. These results suggest that a similar multiscale nature of magnetic loops can be observed on larger scales.

Our simulations results are consistent with the recent observations of small-scale jets with diameter $\sim 200$~km and lifetime of 0.5 -- 4~min \citep{Yurchyshyn2011}, and suggest that the origin of these jets is associated with vertical vortex tubes in the intergranular lanes. Other examples of  fine-structured  loops and upflows are presented by \cite{Ji2012}, who combined BBSO/NST observations in He I 10830~\AA~and TiO 7057~\AA~lines, and found links between the ultrafine loops ($\sim 100$~km wide) and magnetic bright points in the photosphere. In our model such correlation can be explained by the stronger magnetic fields in the vortex tubes, which can efficiently accelerate upflows in the chromosphere.

In addition to the complex and dynamic Stokes I profiles, we found increasing linear polarization during the flow eruptions, which can be detectable with Hinode SP. More detailed studies are required to understand the physical properties and statistical distribution of the small-scale eruptions, and, develop more precise and quantitative spectro-polarimetric diagnostics based on radiative MHD simulations. We plan to investigate the spectropolarimetric properties in different wavelength in more detail, and identify their signatures in the current observational data from SOT/Hinode and ground-based telescopes, such as NST/BBSO, and make also predictions for future space missions, such as Solar-C.


{\it\bf  Acknowledgements.} The simulation results were obtained on the NASA's Pleiades supercomputer at the NASA Ames Research Center. This work
was partially supported by the NASA grants NNX10AC55G and NNH11ZDA001N-LWSCSW, and Oak Ridge Associated Universities.

\bibliographystyle{aa}
.
\end{document}